\documentclass[11pt,fleqn]{article}

\usepackage{a4wide}
\usepackage{cite}
\usepackage{epsf}
\usepackage{rotate}

\renewcommand{\thefootnote}{\fnsymbol{footnote}}    

\newcommand{\beq}{\begin{equation}}
\newcommand{\eeq}{\end{equation}}
\newcommand{\bea}{\begin{eqnarray}}
\newcommand{\eea}{\end{eqnarray}}

\newcommand{\cO}{{\cal O}}
\newcommand{\cE}{{\cal E}}

\begin{document}

\thispagestyle{empty}

\begin{flushright}
TUM-HEP-298/97
hep-ph/9710503 \\
October 1997 \\
\end{flushright}

\begin{center}

\vskip 2cm
\Large
{\bf Anomalous dimensions of four-quark operators \\ in the large $N_f$ limit}
\vskip 1.5cm

\large
Nicolas Pott\footnote[3]{
Supported by the German Bundesministerium f{\"u}r Bildung und Forschung
under contract 06 TM 874 and DFG Project Li 519/2-2.}

\vskip 1cm
\small
{\em Physik-Department, Technische Universit{\"a}t M{\"u}nchen \\
 D-85748 Garching, Germany}
\end{center}

\vskip 2cm
\begin{abstract}
The anomalous dimensions of four-quark operators $(\bar q_i q_j)_{V-A}
(\bar q_k q_l)_{V-A}$ are calculated in the large $N_f$ limit. As
expected, the result is a convergent series without renormalon
ambiguities. Using the approximation of ``Naive Nonabelianization", 
an additional all-order contribution to the anomalous dimension matrix
is obtained
which is somewhat larger than the exact NLO correction itself. Possible
phenomenological applications in nonleptonic $B$ decays and $B^0 -
\bar B^0$ mixing are briefly considered.
\end{abstract}

\renewcommand{\thefootnote}{\arabic{footnote})}

\newpage
\pagenumbering{arabic}

\section{Introduction}

The behaviour of the perturbative series at large orders constitutes
one of the major unsolved problems of quantum field theory. Since it
is in practice impossible to calculate a given physical quantity exactly
to all orders of perturbation theory, the following approach for a
better understanding of the {\em a priori} unkown large-order behaviour
suggests itself: namely, instead of considering all possible diagrams,
one tries to identify certain subclasses of diagrams which become
dominant at sufficiently high orders and can be calculated exactly
to all orders.

Once such a subclass of diagrams is found, it will be useful in three
aspects: (i) for analyzing fundamental questions related to the
summability and/or uniqeness of the perturbative series, as well as
the connection of these problems to non-perturbative issues, (ii) for
estimating the numerical effects of higher order contributions, which
are important in many phenomenological applications, (iii) for having at
one's disposal an independent check of forthcoming exact higher-order
calculations. The latter point should especially be emphasized, since
such calculations will most certainly be performed in a completely
computerized way. Therefore, analytical expressions for a well-defined
part of the full result should be regarded as a helpful and desirable tool
for the testing of these probably very extensive computer programs.

In fact, until now only one of these subclasses of Feynman graphs has
been found : diagrams with an insertion of an arbitrary number of
massless fermion loops (``bubbles") in a gluon line, the so-called
renormalon chains. If a physical quantity $R$ has the perturbative
expansion
\beq R = \sum_{n=0}^{\infty}r_n a^n, \eeq
where $a$ is the
coupling constant of the theory, and if we perform for each
coefficient $r_n$ an additional expansion in $N_f$, the number of
massless fermions in the theory,
\beq r_n=\sum_{m=0}^{n-1}r_n^{(m)}
N_f^m, \qquad n \ge 1, \eeq
 then the coefficients $r_n^{(n-1)}$ will
obviously be determined by diagrams where $n-1$ fermion bubbles
have been inserted into a gluon line of some leading-order graph.

Of course it is doubtful whether these coefficients $r_n^{(n-1)}$
correctly describe the asymptotic behaviour of the perturbative
series or not. Although this will be the case in the limit $N_f
\rightarrow \infty$, at first sight it looks like a gross mistake to
conclude that the real world (with $N_f=5$ in the case of $B$ physics or
$e^+ e^-$ annihilation) resembles at least in
some aspects this fictitious limit. Yet there are some arguments
for such a point of view. Above all they rely on the observation
\cite{lautrup:77,thooft:78} that the amplitude of an individual
fermion-bubble diagram grows in general like $n!$ at high orders. If
one remembers that usually not the amplitude of an individual graph
but only the number of all possible graphs grows with $n!$, and if one
further assumes that all diagrams yield approximately the same
contribution with no preferred sign, then one can argue that the
fermion-bubble diagrams should become dominant at sufficiently high
orders. For recent reviews and
further references see \cite{altarelli:95, fischer:97}.

In practical applications, higher order effects play a significant
role only in QCD with its relative large coupling constant. However,
QCD is exceptionally bad described by the limit $N_f \rightarrow
\infty$, since in this limit asymptotic freedom, one of the most
characteristic features of the strong interaction, is violated
($\beta_0 \sim N_f -33/2$, the first coefficient of the QCD beta
function, becomes positive for $N_f > 16$). As a generalization of the
QED case, where $\beta_0 \sim N_f$, it was therefore proposed
\cite{broadhurst:95,beneke:95,maxwell:95} that in QCD one should expand the
perturbative coefficients $r_n$ in powers of $N_f-33/2$, \beq
r_n=\sum_{m=0}^{n-1}r_n^{[m]} (N_f-33/2)^m, \qquad n \ge 1, \eeq and
that physical quantities are at sufficiently high orders well
described by neglecting all but the coefficients
$r_n^{[n-1]}=r_n^{(n-1)}$ in this expansion. This approximation is
usually referred to as ``Naive Nonabelianization" (NNA).

In the last two years the NNA method was applied to several
observables where a comparison with existing Next-To-Leading-Order
(NLO) or Next-To-Next-To-Leading-Order (NNLO) calculations was
possible. Specifically, analyses were performed of the $R$ ratio and
related quantities \cite{maxwell:95}, of the hadronic decay width
$R_\tau$ of the $\tau$ lepton \cite{ball:95}, of semileptonic B decays
\cite{ball:95a}, and of structure functions in deep inelastic
scattering \cite{stein:96, stein:96a,mankiewicz:97,mankiewicz:97a}. In
all of these, a rather good agreement between NNA and the exact
calculation was claimed at third or even at second order. However, it
should be stressed that a systematic reason for this agreement is
still not known.

It is interesting to extend this program to some other quantities were
exact NLO results are available. To this end, in this paper the
anomalous dimensions of the weak current-current four-quark operators
is calculated in the large $N_f$ limit. Such operators occur in the
effective Hamiltionian for e.\,g.\ weak hadronic decays and
particle-antiparticle-mixing involving mesons. There NLO anomalous
dimension matrix was calculated in \cite{altarelli:81, buras:90},
see also \cite{buchalla:95} for a recent review on the subject of
effective hamiltonians for weak decays.  Clearly, those anomalous
dimensions are no direct physical observables; a complete calculation
of the effective hamiltonian additionally includes the matching of the
effective onto the full theory at the electroweak scale. This
procedure involves the calculation of finite parts of Feynman
diagrams. At a given order in $\alpha_s$, the matching is therefore
considerably more difficult than
the computation of renormalization group functions which is
performed in what follows; nevertheless {\em a posteriori} it turns out
that it is presumably even more important. This point will be discussed
in the last but one section.

The organization of this paper is as follows: In section 2, the scene
is set with the basic definitions of the relevant operators and their
anomalous dimensions; it is then explained how to obtain these anomalous
dimensions from the calculation of truncated Green functions. In
section 3, this calculation is performed and the necessary
renormalization of the gluon field in the large $N_f$ limit is
discussed in detail. From this an analytic large $N_f$ formula for the
anomalous dimensions, summed to all orders in $\alpha_s$, can be
obtained. This is done in section 4. The result is discussed in
section 5, where the NNA approximation is used to estimate the
perturbative contributions beyond next-to-leading order, an estimate
that will be relevant for some phenomenological applications. It is
also discussed how the result for the anomalous dimensions should be
seen in the context of the full large $N_f$ calculation of the
effective Hamiltonian. The paper closes with a short summary in
section 6.  Finally, an appendix is devoted to the so-called
evanescent operators, objects that occur in the course of the
calculation in section 3. In particular, it is proved that these
operators do not contribute to anomalous dimensions in the large $N_f$
limit.

\section{Definitions}

Current-current operators arise due to tree-level $W$ exchange. There
Dirac structure is given by
\beq \label{diracstructure}
\hat \gamma  = {\bar q}_i \gamma^\mu (1-\gamma_5) q_j
\, {\bar q}_k \gamma^\mu (1-\gamma_5) q_l,
\eeq
where $i,j,k,l$ denote flavour indices. Due to exchange of gluons
between the quark legs, two different color
structures can arise, in symbolic form written as
\beq
{\bf 1} = \delta_{\alpha \beta} \, \delta_{\gamma \delta}, \qquad
{\bf \tilde 1} = \delta_{\alpha \delta} \, \delta_{\beta \gamma},
\eeq
where $\alpha,\beta,\gamma,\delta$ denote the color indices of the
first, second, third and fourth quark field $q$ in Eq.\ (\ref{diracstructure}).
In this compact notation, the two operators we are dealing with are
given by
\bea \label{2opbasis}
{\cal  O}_1 &=& \hat \gamma {\bf \tilde 1}, \nonumber \\
{\cal  O}_2 &=& \hat \gamma {\bf 1}.
\eea
For further purposes we also introduce the linear combinations
\beq \label{2diagbasis}
{\cal O}_{\pm} = {1 \over 2} ({\cal O}_2 \pm {\cal O}_1).
\eeq
In this basis, the anomalous dimension matrix will be
diagonal. If we now calculate some truncated Green function $\langle {\cal O}_i
\rangle$ with an insertion of one of these operators, the result will
in general be divergent even after renormalization of fields and QCD
coupling. In order to
achieve a finite result, we make use of our freedom to redefine the
operator basis\footnote{In general, one has  to include additional
``evanescent" operators in the operator basis, since they unavoidably
arise while calculating loop diagrams like those of
Fig.\ 1. Accordingly, these evanescent operators also participate in
the multiplicative renormalization of Eq.\ (\ref{oprenom}). However, in
the large $N_f$ limit they can safely be neglected, as shown
in the appendix.} according to
\beq \label{oprenom}
{\cal O}_i = Z_{ij} {\cal O}_j^R.
\eeq
If we additionally renormalize the four external fields,
\beq
q = Z_\psi^{1/2}q^R,
\eeq
then we can write the truncated Green function as
\beq \label{greenfunc}
\langle {\cal O}_i
\rangle = Z_\psi^{-2} Z_{ij} \langle {\cal O}_j^R \rangle,
\eeq
where the $\langle {\cal O}_j^R \rangle$'s are now finite. The
subtractions will be performed in the $MS$ scheme throughout, so
(working in $D=4-2 \epsilon$ dimensions) the
renormalization constants $Z_{ij}$ will have the form
\beq \label{zij}
Z_{ij}=1+\sum_{k=1}^{\infty} {Z_{ij}^{(k)} \over \epsilon^k}.
\eeq
$Z_{ij}$ is a function of the renormalized coupling and therefore depends
on the renormalization scale $\mu$. The anomalous dimension
matrix of the operators ${\cal O}_i$ is defined as
\beq \label{anomdim}
\gamma_{ij} = (Z^{-1})_{ik} \, \mu {d Z_{kj} \over d \mu}. 
\eeq
Introducing the quantity
\beq
\tilde Z_{ij} = Z_\psi^{-2} Z_{ij},
\eeq
which can, according to Eq. (\ref{greenfunc}), be read off directly from
the divergent pieces of the
truncated Green functions $\langle {\cal O}_i \rangle$, we may write
Eq.\ (\ref{anomdim}) as 
\beq \label{2anomdimmaster}
\gamma_{ij} = \tilde \gamma_{ij} + 2 \gamma_{\psi},
\eeq
with
\beq
\tilde \gamma_{ij} = (\tilde Z^{-1})_{ik} \, \mu {d \tilde Z_{kj} \over
 d \mu}, \qquad \gamma_{\psi}=Z_\psi^{-1} \mu {d Z_\psi \over d \mu}.
\eeq
Using the equivalents of Eq.\ (\ref{zij}) for $\tilde Z_{ij}$ and
$Z_\psi$, the chain rule $d/d \mu = (dg/d \mu)\,d/dg$ (where $g$ is the
renormalized QCD coupling) and the renormalization group equation for
$g$ one obtains \cite{floratos:77}
\beq \label{2andim}
\tilde \gamma_{ij}= -2 g^2 {d \tilde Z_{ij}^{(1)} \over g^2}, \qquad
\gamma_\psi = -2 g^2 {d Z_\psi^{(1)} \over dg^2},
\eeq
i.\,e.\ the anomalous dimensions are directly related to the
$1/\epsilon$ pole part of the corresponding renormalization constant.

\section{Calculation of Green functions with bubble insertions}

We now proceed to calculate the truncated Green functions $\langle
{\cal O}_1 \rangle$ and $\langle {\cal O}_2 \rangle$ with an insertion
of $n$ fermion bubbles into the gluon line, i.\,e.\ the diagrams shown
in Fig.\ 1.
\begin{figure}[thb]
\centerline{
\epsfysize=11cm
 \rotate[r]{
 \epsffile{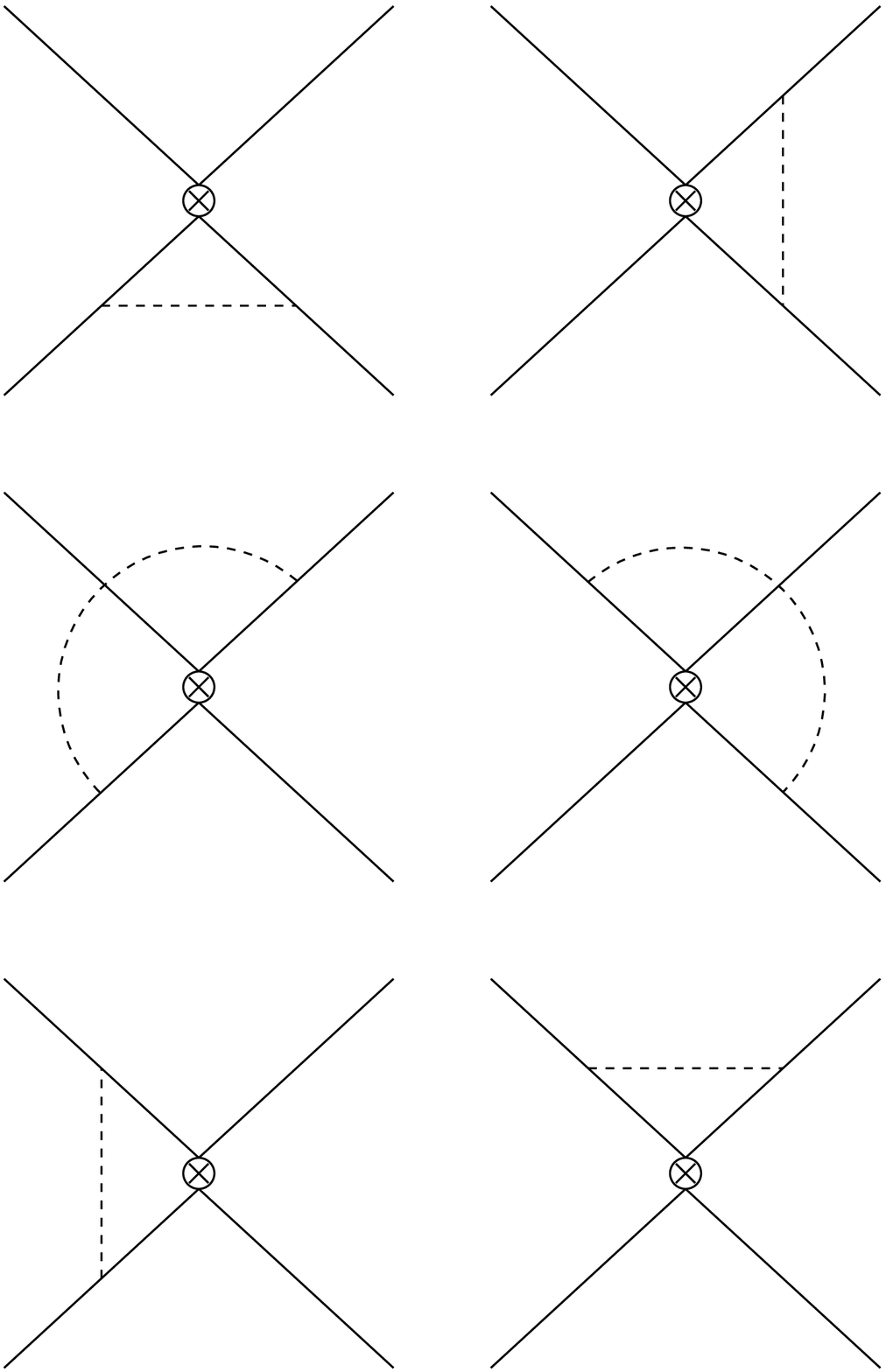}}
}
\caption{\small The six Feynman graphs relevant for the anomalous
dimensions of the operators $\cO_1$ and $\cO_2$. The dashed line
denotes the gluon propagator with an insertion of $n$ fermion bubbles
and the appropriate counterterms.}
\label{fig1}
\end{figure}
In order to determine the quark wave function anomalous
dimension $\gamma_\psi$, we will additionally calculate the truncated
two point function depicted in Fig.\ 2.
\begin{figure}[thb]
\centerline{
\epsfysize=6cm
\rotate[r]{
\epsffile{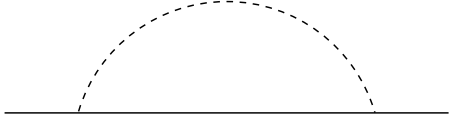}}
}
\caption{\small The Feynman graph relevant for the anomalous
dimension of the quark wave function. The dashed line
again denotes the gluon propagator with an insertion of $n$ fermion bubbles
and the appropriate counterterms.}
\label{fig2}
\end{figure}
To this end, we need as  a first step an expression for the gluon propagator
with an insertion of $n \ge 0$ fermion bubbles (but no counterterms). A
straight forward calculation in $D=4-2 \epsilon$ dimension yields
\beq \label{3gluonprop}
\Pi_{\mu \nu}^{(n)}(k) = {-i \over k^2} \left ( g_{\mu \nu} - {k_\mu
 k_\nu \over k^2} \right ) \left [ - \Pi(k^2) \right ]^n.
\eeq
Here, the function $\Pi(k^2)$ stems from the one loop calculation of
the fermion bubble,
\beq \label{3fermionbub}
\Pi(k^2)= {\alpha_s \over \pi} N_f \left ( {4 \pi \mu^2 \over -k^2}
\right )^\epsilon \Gamma(\epsilon) \, \mbox{B}(2-\epsilon,2-\epsilon),
\eeq
and $k$ is the momentum flowing through the propagator. It should be
stressed that Eq.\ (\ref{3gluonprop}) is valid in any gauge for $n \ge
1$. Only for $n=0$ it is the specific result of the Landau gauge. So
the fermion bubble contribution is for itself a gauge independent
quantity, as of course must be the case.

Now one can insert the effective propagator (\ref{3gluonprop}) into
the diagrams of Fig.\ 1 and perform the loop integrations via
introduction of Feynman parameters as usual. Adding the contributions
from all six diagrams, we find
\beq \label{3master1}
\left ( \begin{array}{c}  \langle {\cal O}_1 \rangle^{bare} \\
 \langle {\cal O}_2 \rangle^{bare} \end{array} \right ) =
\left \{ 1- {\alpha_s \over 4 \pi} \left ({\alpha_s N_f \over 6 \pi} 
\right )^n
{(-1)^n
\over (n+1) \epsilon^{n+1}} \, \mbox{F}(\epsilon,(n+1) \epsilon) \, M_{c}
\right \}  \left ( \begin{array}{c}  \langle {\cal O}_1 \rangle^{tree} \\
 \langle {\cal O}_2 \rangle^{tree} \end{array} \right ),
\eeq
with
\beq \label{3master2}
F(\epsilon,z)=\left ( {4 \pi \mu^2 \over -\lambda^2} \right ) ^z
\Gamma(1+z) \, \Gamma(1-z) \Bigl [ 6 \epsilon \, \Gamma(\epsilon) \,
\mbox{B}(2-\epsilon, 2-\epsilon) \Bigr ]^{z/\epsilon-1}
{6-4 \epsilon \over \Gamma(3-\epsilon)},
\eeq
\beq \label{3master3}
M_{c} = \left ( \begin{array}{cc}  - {1 \over N}+\epsilon \left ( C_F-{1
\over N} \right ) & 1+\epsilon \\ 1+\epsilon &  - {1 \over
N}+\epsilon  \left ( C_F-{1 \over N} \right ) \end{array} \right )
\eeq
and $C_F=(N^2-1)/(2N)$, $N=3$. $\langle {\cal O}_i \rangle^{tree}$ denote
the trivial tree-level matrix elements of the operators ${\cal O}_i$,
and the auxiliary function $\mbox{F}(\epsilon,z)$ has been defined in
such a way such $\mbox{F}(0,0)$, $\mbox{F}(\epsilon,0)$ and
$\mbox{F}(0,z)$ are finite. From a calculational point of view,
Eqs.\ (\ref{3master1}) -- (\ref{3master3}) constitute already the main
result of this paper, so several remarks are in order:

(i) The index ``bare''  indicates that until now no renormalization has
been performed. In the next step we will renormalize the gluon field
so that the transition  $\langle {\cal O}_i \rangle^{bare} \rightarrow
\langle {\cal O}_i \rangle$ is achieved. The final operator
renormalization $\langle {\cal O}_i \rangle \rightarrow \langle {\cal
O}_i^R \rangle$, cf.\ Eq.\ (\ref{oprenom}), is the subject of the following
section.

(ii) In calculating Eq.\ (\ref{3master1}) we set all external momenta
to zero which is allowed since the renormalization constants we are
interested in do not depend on the external states. Infrared divergent
integrals were regulated by introducing a finite gluon mass $\lambda$
that appears as the only scale (apart from the renormalization scale
$\mu$) in Eq.\ (\ref{3master2}).

(iii) In order to obtain Eq.\ (\ref{3master1}) we had to project complicated
Dirac structures, generically of the form $\Gamma \otimes \tilde
\Gamma$, onto the simpler $V-A \otimes V-A$ structure of the operators
${\cal O}_i$, i.\,e.\ we used identities like
\beq \label{3projection}
\gamma^\mu \gamma^\nu \gamma^\lambda(1-\gamma_5) \otimes \gamma_\mu 
\gamma_\nu \gamma_\lambda (1-\gamma_5) =
4 \, (4-\epsilon-\epsilon^2) \, \gamma^\mu (1-\gamma_5) \otimes \gamma_\mu 
(1-\gamma_5).
\eeq
A general method for obtaining such relations is explained e.\,g.\ in 
\cite{herrlich:95}. However, such a projection is a very peculiar
procedure since it has to be done in $D=4-2 \epsilon$ dimensions
where, strictly speaking, no complete finite basis $\{\gamma^{(i)} \otimes
\tilde \gamma^{(i)} \} $ of Dirac structures on which to project can
be given. Eq.\ (\ref{3projection}) should therefore be understood as
the projection on a certain subset of such a basis, namely, the one
that forms the complete basis in $D=4$ dimensions. Only in this sense
the coefficient $4\,(4-\epsilon-\epsilon^2)$ in Eq.\ (\ref{3projection}) is
a well defined quantity. Rigorously, the discrepancy between the
l.h.s. and the r.h.s. of Eq.\ (\ref{3projection}) defines an
evanescent operator (vanishing in the limit $\epsilon \rightarrow
0$) which has to be treated correctly in higher order calculations
\cite{herrlich:95}. However, it is shown in the appendix that such
operators are irrelevant for the purpose of this paper.

In order to renormalize the gluon field, the quite ingenious
method described by Beneke and Braun in the appendix of 
\cite{beneke:94} is used. One first needs the counterterm appropriate to
cancel the divergence of one fermion bubble. From Eq.\
(\ref{3fermionbub}) it can be read off to be ${\alpha_s \over 6 \pi}
N_f {1 \over \epsilon}$ in the $MS$ scheme. Since there are
$n!/(k!\,(n-k)!)$ possibilities to replace $k$ out of $n$ fermion
bubbles with this counterterm, and since all these possibilities have
to be added, one can infer from Eq.\ (\ref{3master1}) the following
expression for the (partially) renormalized Green functions: \bea
\label{3masterren} \left ( \begin{array}{c} \langle {\cal O}_1 \rangle
\\ \langle {\cal O}_2 \rangle \end{array} \right ) &=& \Biggl \{
\Biggr.  1- {\alpha_s \over 4 \pi} \left ({\alpha_s N_f \over 6 \pi} 
\right )^n {1 \over \epsilon^{n+1}} \nonumber \\ &&
\Biggl. \sum_{k=0}^n \left [ \left (
\begin{array}{c} n \\ k \end{array} \right )
{(-1)^k
\over k+1 } \mbox{F}(\epsilon,(k+1) \epsilon) \right ] M_{c}
\Biggr \}  \left ( \begin{array}{c}  \langle {\cal O}_1 \rangle^{tree} \\
 \langle {\cal O}_2 \rangle^{tree} \end{array} \right ).
\eea
We called this the partially renormalized Green functions, because
operator and wave function renormalization are still to be performed.

It is gratifying that the finite sum in Eq.\ (\ref{3masterren}) can
be solved. Expanding $\mbox{F}(\epsilon,z)$ in powers of
$z$,
\beq
\mbox{F}(\epsilon,z) = \sum_{j=0}^{\infty} f_j(\epsilon) \, z^j,
\eeq
we have
\beq \label{3aux1}
\sum_{k=0}^n \left [ \left ( \begin{array}{c} n \\ k \end{array}
\right ) {(-1)^k \over k+1} \mbox{F}(\epsilon,(k+1) \epsilon) \right ]
= \sum_{j=0}^{n+1} f_j(\epsilon)\, \epsilon^j \left [
\sum_{k=0}^n   \left ( \begin{array}{c} n \\ k \end{array}
\right ) (-1)^k (k+1)^{j-1} \right ].
\eeq
The sum over $j$ stops at $j=n+1$, since the remaining terms do not
contribute in the limit $\epsilon \rightarrow 0$ (note that Eq.\
(\ref{3aux1}) is still to be multiplied with an overall factor 
$1/\epsilon^{n+1}$). Using appropriate formulas from 
\cite{hansen:75,gradshteyn:65}, the sum over $k$ in Eq.\ (\ref{3aux1})
can easily be performed. It is different from zero only in the cases
$j=0$ and $j=n+1$,
\beq \label{3aux2}
\sum_{k=0}^n \left [ \left ( \begin{array}{c} n \\ k \end{array}
\right ) {(-1)^k \over k+1} \mbox{F}(\epsilon,(k+1) \epsilon) \right ]
= f_0(\epsilon) {1 \over n+1} + f_{n+1} \, (\epsilon) \epsilon^{n+1}
(-1)^n n! \; \; .
\eeq
The term proportional to $\epsilon^{n+1}$ can be discarded since it
yields only a finite contribution in the limit $\epsilon \rightarrow 0$
which is irrelevant for the anomalous dimensions. With
\beq
f_0(\epsilon) = {3-2 \epsilon \over 3 \Gamma(1+\epsilon)
\mbox{B}(2-\epsilon,2-\epsilon) \Gamma(3-\epsilon)}
\eeq
from Eq.\ (\ref{3master2}), one finally obtains
\bea \label{3masterop}
\left ( \begin{array}{c}  \langle {\cal O}_1 \rangle \\
 \langle {\cal O}_2 \rangle \end{array} \right ) &=&
\biggl \{ 1- {\alpha_s \over 4 \pi} \left ({\alpha_s N_f \over 6 \pi} 
\right )^n  
\biggl [ {1 \over (n+1) \epsilon^{n+1}} \biggr. \biggr.
\nonumber \\ && \biggl. \biggl. {3-2 \epsilon \over 3 \Gamma(1+\epsilon)
\mbox{B}(2-\epsilon,2-\epsilon) \Gamma(3-\epsilon)} + \mbox{finite}
 \biggr ] M_{c}
\biggr \}  \left ( \begin{array}{c}  \langle {\cal O}_1 \rangle^{tree} \\
 \langle {\cal O}_2 \rangle^{tree} \end{array} \right ).
\eea
From this formula the renormalization constants $\tilde Z_{ij}$ can
immediately be read off.

A completly analogous calculation can be done for the fermion self
energy with an insertion of $n$ fermion bubbles, see Fig.\ 2. This
calculation yields
\bea \label{3masterwave}
i\Sigma^{(n)}(p)&=& i p \hspace{-0.5em}/ \hspace{0.1cm} \biggl \{ 1 + {\alpha_s \over 4 \pi} \left ( {\alpha_s
N_f \over 6 \pi} \right )^n \biggr. \nonumber \\ 
& & \hspace{0.5cm} \biggl. \left [{ 1 \over (n+1) \epsilon^{n+1}}
{\epsilon   (3-2 \epsilon) \Gamma(4-2 \epsilon)
\over 6 \Gamma(1+e) \Gamma(2-\epsilon)^2 \Gamma(3-\epsilon) } +
\mbox{finite}
 \right ] C_F
\biggr \},
\eea
from which the wave function renormalization constant $Z_\psi$ can be
read off. Eq.\ (\ref{3masterwave}) is again gauge independent for $n
\ge 1$ and specific to Landau gauge in the case $n=0$.

\section{Extraction of anomalous dimensions}

Summarizing the results of the previous section, renormalization
constants in the large $N_f$ limit may in general be written as
\beq \label{41}
Z=1-\sum_{k=0}^n {\alpha_s \over 4 \pi} \left ( {\alpha_s N_f \over 6
\pi} \right ) ^k {f_0(\epsilon) \over (k+1) \epsilon^{k+1}} 
\eeq
at any {\em fixed} order $n+1$ in perturbation theory. Here,
$f_0(\epsilon)$ is some function of $\epsilon$ which may also contain
color factors. However, we prefer to have at our disposal 
an {\em all-order} result for the corresponding anomalous dimension.
This can be achievied by expanding $f_0(\epsilon)$ in powers of
$\epsilon$,
\beq \label{42}
f_0(\epsilon)=\sum_{j=0}^\infty \tilde f_j \, \epsilon^j.
\eeq
Inserting this into Eq.\ (\ref{41}), we obtain for the $1/\epsilon$
part of the renormalization constant, in the notation of Eq.\
(\ref{zij}),
\beq
Z^{(1)} = - \sum_{k=0}^n {\alpha_s \over 4 \pi} \left ( {\alpha_s N_f
\over 6 \pi} \right )^k {\tilde f_k \over k+1},
\eeq
which can formally be summed up to all orders by setting $n$ to infinity,
\beq
Z^{(1)}_{all-orders}= - \sum_{k=0}^\infty {\alpha_s \over 4 \pi} \left ( {\alpha_s N_f
\over 6 \pi} \right )^k {\tilde f_k \over k+1}.
\eeq
From this the corresponding anomalous dimension is obtained according
to Eq.\ (\ref{2andim}) as
\bea
\gamma_{all-orders} &=& -2 \alpha_s {d Z^{(1)}_{all-orders} \over d
\alpha_s} \nonumber \\
&=& {\alpha_s \over 2 \pi} \sum_{k=0}^{\infty} \left ( {\alpha_s N_f
\over 6 \pi} \right )^k \tilde f_k \nonumber \\
&=& {\alpha_s \over 2 \pi} \, f_0 \left ({\alpha_s N_f \over 6 \pi}
\right),
\eea
where in the last step Eq.\ (\ref{42}) was employed.

Thus we get from Eq.\ (\ref{3masterwave}) for the
anomalous dimension of the quark field
\beq
\gamma_\psi = {\alpha_s \over 2 \pi} C_F \, f_\psi \left ( \alpha_s N_f
\over 6 \pi \right),
\eeq
with
\beq
f_\psi(x) = -{1 \over 6} x  (3-2 x) {\Gamma(4-2 x) 
\over \Gamma(1+x) \Gamma(2-x)^2 \Gamma(3-x)},
\eeq
in agreement with a result that Gracey \cite{gracey:93} derived some
years ago using a completely different method.
Having found this, we finally obtain from Eqs.\ (\ref{2anomdimmaster}) and
(\ref{3masterop}) the following formula for the anomalous dimensions of
the operators ${\cal O}_1$ and ${\cal O}_2$ in the large $N_f$ limit:
\beq
\gamma_{ij} = \tilde \gamma_{ij} + 2 \gamma_\psi = {\alpha_s \over 2
\pi} \,
f_{\cal O} \left ( {\alpha_s N_f \over 6 \pi} \right ) \left (
\begin{array}{cc} -1/N & 1 \\ 1 & -1/N \end{array} \right ),
\eeq with \beq f_{\cal O}(x)= {1 \over 3} (1+x) (3-2 x) {\Gamma(4-2 x)
\over \Gamma(1+x) \Gamma(2-x)^2 \Gamma(3-x)}.  \eeq This can
alternatively be written in the diagonal operator basis
(\ref{2diagbasis}), where one has \beq \label{4masterop} \gamma_{\pm}
= \pm {\alpha_s \over 2 \pi} \, f_{\cal O} \left ( {\alpha_s N_f \over 6
\pi} \right ) {N \mp 1 \over N}.  \eeq
Reexpanding in powers of
$\alpha_s$, our result (\ref{4masterop}) agrees with the
 $\alpha_s^2 N_f$ part of the exact NLO result as given in Eq.\ (5.2) of 
\cite{buras:90}. One should mention that although our findings were
derived in the $MS$ scheme, they are of course also valid in the
 $\overline{MS}$
scheme (with the replacement $\alpha_s^{MS} \rightarrow
\alpha_s^{\overline {MS}}$ understood). This is because the $MS$
scheme is related to the latter simply by a multiplicative scale
redefinition $\mu \rightarrow (e^{\gamma_E}/4 \pi)^{1/2} \mu$, so that
the functional dependencies of renormalization group functions on
$\alpha_s$ do not change. On the other hand, it should
be emphasized that the result
for the anomalous dimensions $\gamma_{\pm}$ in general as well as in
the large $N_f$ limit depends on the regularization scheme
used for $\gamma_5$ in $D$ dimensions: all our
results were obtained in the ``Naive Dimensional Regularization"'
(NDR) scheme with a naively anticommuting $\gamma_5$.

\section{Discussion of results}

As explained in the introduction, the large $N_f$ limit of QCD
unfortunately corresponds by no means to the physical world.
Nevertheless it
is widely conjectured that one can still make some contact to reality,
if one uses the results of the large $N_f$ limit as a starting point
for the so-called NNA approximation, i.\,e.\ if one replaces $N_f$ by
$N_f-33/2$ in these results. This procedure, motivated by the analogy
to QED where $N_f$ is essentially the first coefficient of the QED
beta function, may also be called the large $\beta_0$
approximation. With $\beta_0= 2N_f/3-11$, it amounts to the
replacement
\beq \label{51} {\alpha_s N_f \over 6 \pi} \rightarrow {\alpha_s
\beta_0 \over4 \pi}
 \eeq
 in Eq.\ (\ref{4masterop}).
To the knowledge of the author, the only justification for this
procedure is a certain phenomenological success; no further comments
on its validity will therefore be found in this paper. Instead, 
 the substitution (\ref{51}) will be used as an
 admittedly crude and naive prescription for
estimating higher-order effects.

Before doing so it is interesting to investigate how well the NNA
approximation works already at NLO, where a comparison with the exact result
is possible.  Expanding
\beq 
\gamma_\pm = \gamma^{(0)}_\pm \, {\alpha_s \over 4 \pi} +
\gamma^{(1)}_\pm
\left ( {\alpha_s \over 4 \pi} \right )^2 + \ldots,
\eeq
the exact result \cite{buras:90} in the NDR scheme reads
\beq \label{55}
\gamma_+^{(1)} = {2 \over 3} \beta_0 + {1 \over 3}, \qquad
\gamma_-^{(1)} = - {4 \over 3} \beta_0 -{86 \over 3}.
\eeq
In the NNA approach, the constant terms are neglected. With $N_f=5$ or
$\beta_0=-23/3$, this is obviously a very good (accurate to 7\%)
approximation for $\gamma_+^{(1)}$, but also a very bad (wrong sign)
approximation for $\gamma_-^{(1)}$. So one cannot claim from this
comparison that the NNA approximation generally works
well at low orders\footnote{See, however, the remark at the end of
this section.}.

Despite this somewhat discouraging observation, let us now ask for the
contributions of all orders beyond NLO in the NNA approach. Evaluating
the anomalous dimensions at a scale $\mu \simeq 5\, \mbox{GeV}$
or $\alpha_s(\mu) \simeq 0.21$, we have
\bea \label{5adnumerics}
\gamma_+ &=& +0.0670\;\mbox{(LO)} - 0.0014 \;\mbox{(exact NLO)}  - 0.0021 \nonumber, \\
\gamma_- &=& -0.1350\;\mbox{(LO)}- 0.0053 \;\mbox{(exact NLO)} + 0.0043.
\eea
Here, the first two numbers are exact result and the third number
indicates the summed contribution from all orders beyond NLO in the NNA
approximation (i.\,e.\ the terms $\alpha_s^3 \beta_0^2$, $\alpha_s^4
\beta_0^3,\ldots$), obtained from Eq.\ (\ref{4masterop}). We notice that
this contribution is of the same order of magnitude as the exact NLO
contribution itself. 

One can go one step further and examine to what extent the Wilson
coefficients $C_i(\mu)$ of the operators $\cO_i$ are affected by these
additional fermion-bubble contributions to the anomalous
dimensions. We will investigate here two cases that are
of some phenomenological interest: nonleptonic $B$ decays and $B^0 -
\bar B^0$ mixing.

\medskip
\noindent
{\em (i) Nonleptonic B decays}

\medskip \noindent
Only decays of the $b$ quark into three different flavours,
e.\,g.\ $b \rightarrow c \bar u d$, are considered.
For these decay modes, no penguin
operators can occur and the complete effective Hamiltonian is given by
\beq
{\cal H}_{eff} = {G_F \over \sqrt{2}} \, V_{CKM} \, [C_1(\mu) \cO_1 +C_2(\mu)
\cO_2],
\eeq
with the four-quark operators $\cO_i$ as defined in Eq.\ (\ref{2opbasis}).
The renormalization group analysis is most easily be done in the
diagonal basis $\cO_\pm$ with the corresponding Wilson coefficients
$z_\pm=C_2 \pm C_1$. These Wilson coefficients obey the
renormalization group equation
\beq \label{5rge}
\mu {d \over d \mu} z_\pm(\mu) = \gamma_\pm z_\pm(\mu)
\eeq
which can be solved numerically using the anomalous dimension
(\ref{4masterop}) and the initial condition \cite{buras:90}
\beq \label{5nlbmatching}
z_+(M_W) = 1+{11 \over 3} {\alpha_s(M_W) \over 4 \pi},\qquad
z_-(M_W) = 1-{22 \over 3} {\alpha_s(M_W) \over 4 \pi}.
\eeq
In this way we obtain (after transforming back to the basis $\cO_{1,2}$
which is more commonly used) at $\mu=m_b \simeq 5 \, \mbox{GeV}$:
\bea \label{5nlbmaster}
C_1(m_b) &=& +1.0952 \; \mbox{(LO)} -0.0250 \; \mbox{(exact NLO)} - 0.0017,
\nonumber \\
C_2(m_b) &=& -0.2250 \; \mbox{(LO)} +0.0575 \; \mbox{(exact NLO)} +
0.0043.
\eea
These equations should be read in analogy to Eq.\ (\ref{55}): the first
two numbers are the exact LO and NLO results, the last number
indicates the summed contribution beyond NLO from the anomalous
dimensions in the NNA approximation\footnote{In the NNA approximation
(essentially corresponding to the large $\beta_0$ limit of QCD) one
should for consistency always use the
{\em LO} running coupling $\alpha_s(\mu)$, neglecting higher order
corrections to the QCD beta function. Both of the last numbers in Eq.\
(\ref{5nlbmaster}) were nevertheless obtained by solving Eq.\
(\ref{5rge}) with the {\em NLO} running
coupling (and the anomalous dimensions exact up to NLO), since we 
wished to estimate the sum of all
contributions {\em beyond NLO} in the NNA approximation.}. 
We conclude that this
contribution is completely negligible as compared to the NLO
correction. The reason for this is that the NLO contribution itself is
dominated by the large matching correction (\ref{5nlbmatching}) so that 
the small modifications due to (\ref{5adnumerics}) in the running 
become unimportant.

\medskip
\noindent
{\em (ii) $B^0-\bar B^0$ mixing}
\medskip

\noindent
The effective Hamiltonian for $B^0 - \bar B^0$ - mixing is \cite{buras:90a}
\beq
{\cal H}_{eff}={G_F^2 \over 16 \pi^2}\, M_W^2\, \vert V_{tb}^\ast
V_{td} \vert^2 \, z_+^{B \bar B}(\mu)\cO_+.
\eeq
It contains only the operator $\cO_+$ for which the NNA approximation
worked very well even at NLO. This leads to the conjecture that NNA
estimates of higher orders might be to some extent reliable here.
The initial condition for the Wilson coefficient 
$z_+^{B \bar B}(\mu)$ is a complicated function of the
top mass which can be found in the appendix of
\cite{buras:90a}. Evaluating this function at $m_t(M_W) \simeq 177 \,
\mbox{GeV}$, we obtain
\beq
z_+^{B \bar B}(M_W) = 2.624 + 2.798 {\alpha_s(M_W) \over 4 \pi}.
\eeq
The evolution down to the low scale with Eq.\ (\ref{5rge}) then yields 
\beq \label{5bbmaster}
z_+^{B \bar B}(m_b)  = 2.284 \; \mbox{(LO)} + 0.028 \; \mbox{(exact NLO)} + 0.006.
\eeq
Again, the summed contributions beyond NLO are minute and can safely
be neglected.

From Eqs.\ ({\ref{5nlbmaster}) and (\ref{5bbmaster}) we conclude that
the higher-order fermion-bubble contributions to the anomalous
dimensions $\gamma_{\pm}$ are irrelevant in phenomenological
applications. The reason for this is not the generic smallness of
these additional contributions, cf.\ Eq.\ (\ref{5adnumerics}), but the
smallness of the running in the current-current sector in general.
Our analysis indicates that the
termination of the perturbative series at NLO is in this case a
very good approximation, with an associated theoretical error of
almost zero.

However, a word of caution is in order here. We examined in this paper
only the large $N_f$ contribution to the {\em anomalous dimension}. The
result Eq.\ ({\ref{4masterop}) was a perfectly convergent series, essentially a product of
gamma functions with $\alpha_s N_f/6 \pi$ as argument, without any
renormalon ambiguity. This is in fact the general structure of
renormalization group functions in the large $N_f$ limit
\cite{mankiewicz:97a, beneke:94, gracey:93}. Conversely, this implies that
there is no $n!$ growth of the perturbative coefficients and so there is
no longer a good reason to believe that at sufficiently high orders
the large $N_f$ contributions become dominant. In the best case we can
consider them as a guess of what {\em may} happen at higher orders.

The $n!$ growth is indeed not present (at least in dimensional
regularization) in the divergent parts of Feynman diagrams but only
in the {\em finite} parts - precisely the pieces we neglected in this
calculation. This somewhat surprising statement can be verified
directly from Eq.\ (\ref{3aux2}). The finite parts manifest themselves
in corrections to the matching of the effective onto the full theory,
i.\,e.\ they contribute to the initial conditions $C_i(M_W)$ of the
Wilson coefficients at the electroweak scale. So this initial
condition is the place where one really expects renormalon singularities in
the Borel plane. Hence, a complete large $N_f$ analysis should
certainly include these matching corrections -- a task which remains to
be done in the future. If the general perturbative expansion of a
Wilson coefficient is written as
\beq
C(\mu)=\sum_{n=0}^\infty \sum_{m=0}^n C_{nm} \, 
\alpha_s^n \left ( \ln {\mu \over
M_W} \right )^m,
\eeq
with expansion coefficients that are by themselves power series in $N_f$,
\beq
C_{nm}=\sum_{k=0}^{n-1} C_{nm}^{(k)} N_f^k,
\eeq
then the calculation of the anomalous dimension provides us, in
principle, with all the coefficients $C_{nm}^{n-1}$ with $m>0$, i.\,e.\
with all the logarithmic pieces. The matching calculation on the other
hand gives us the coefficients $C_{n0}^{n-1}$ which will grow as
$(n-1)!$ and will therefore presumably dominate the whole perturbative
expansion at sufficiently high orders. Strictly speaking, this means
that the whole renormalization group evolution from $\mu \simeq M_W$
down to $\mu \simeq m_b$, i.\,e.\ the summation of large
logarithms, becomes meaningless in the large  $N_f$ limit, since the
perturbative series in this limit is (again: at sufficiently high
orders) dominated by the constant and not
by the logarithmic pieces, even if the latter are large. More work
needs to be done in order to clarify these questions completely.

\section{Summary}

In this paper the anomalous dimensions of the current-current
four-quark operators $\cO_1$ and $\cO_2$ were calculated in the large
$N_f$ limit in the $MS$ (or $\overline{MS}$) scheme. The calculation
was done via direct computation of Feynman diagrams in dimensional
regularization, with a special emphasis on the role of evanescent
operators. Our findings agree with the $\alpha_s^2 N_f$ part of the
exact NLO result and may furthermore serve as an independent check for
future NNLO calculations. Using the NNA approximation, we estimated
with our result the total contribution of all orders beyond NLO to the
anomalous dimension. This contribution was found to be of the same
order of magnitude as the exact NLO correction itself. We also
investigated to which extent this contribution affects physical
quantities. As an example, we considered the Wilson coefficients
relevant for nonleptonic $B$ decays and $B^0-\bar B^0$ mixing. In
these two cases, the effects were negligibly small and we concluded
that the truncation of the perturbative series at NLO is a very good
approximation in the current-current sector. Finally, future steps
were discussed that will be necessary for a complete large $N_f$
analysis of effective weak Hamiltonians.

\section*{Acknowledgments}
I would like to thank Manfred M{\"u}nz for helpful discusions concerning
evanescent operators and Andrzej Buras for critically reading the
manuscript.

\begin{appendix}
\section*{Appendix: Evanescent operators}

As we observed in Sec.\ 3, evanescent structures as e.\,g.
\beq \label{app1}
\cE_1=\gamma^\mu \gamma^\nu \gamma^\lambda(1-\gamma_5) \otimes \gamma_\mu 
\gamma_\nu \gamma_\lambda (1-\gamma_5) -
4 \, (4-\epsilon-\epsilon^2) \, \gamma^\mu (1-\gamma_5) \otimes \gamma_\mu 
(1-\gamma_5)
\eeq 
unavoidably arise while calculating matrix elements of weak four-quark
operators\footnote{It is essential for the following argument to
define the evanescent operators in such a way that $\cE_i \perp \cO$ in
$D=4-2 \epsilon$ dimensions so that the coefficients $c_i$ in
Eq.\ (\ref{app3}) start at $O(\alpha_s)$ and not at $O(\alpha_s^0)$. This
condition fixes the coefficient $4 \, (4-\epsilon-\epsilon^2)$ in Eq.\
(\ref{app1}) uniquely. In general, however, the $O(\epsilon)$ part of
this coefficient can be chosen arbitrarily (implying different
definitions of the evanescent opertors) without affecting the physical
results. For a detailed discussion on the subject of
evanescent operators we refer to the work of Herrlich and Nierste
\cite{herrlich:95}.}. For a consistent calculation these operators must be
included from the very beginning,, i.\,e.\ one should start such a calculation
from a ``true'' effctive Hamiltonian
\beq \label{app2}
{\cal H}_{eff} = C_\cO \cO + \sum_{j=1}^{\infty} C_{\cE_j} \cE_j,
\eeq
where the second term constitutes an infinite tower of evanescent
operators vanishing in the limit $\epsilon \rightarrow 0$.
To avoid an embarrassing amount of indices, here and in the following
two simplifying assumptions are made: (i) we are dealing with only one
physical operator $\cO$, (ii) only one new evanescent
operator arises at each loop, i.\,e.\ the $n$ loop matrix elements can be
written as
\bea \label{app3}
\langle \cO \rangle^{(n)} &=& a \langle \cO \rangle^{(0)} + \sum_{j=1}^n
b_j \langle \cE_j \rangle^{(0)}, \nonumber \\
\langle \cE_i \rangle^{(n)} &=& c_i \langle \cO \rangle^{(0)} +
\sum_{j=1}^\infty d_{ij} \langle \cE_j \rangle^{(0)},
\eea
where the sum in the first equation stops at $n$ and not at infinity,
and with some coefficients $a$, $b_i$, $c_i$, $d_{ij}$ which are
computed by explicit calculation. Of course both assumptions can 
easily be relaxed.

Now the crucial point is that we wish to avoid finite contributions
from evanescent operators to the physical matrix element
\beq \label{app4}
\langle {\cal H}_{eff}^R \rangle =
 C_\cO \langle \cO^R \rangle + \sum_{j=1}^{\infty} C_{\cE_j} \langle
\cE_j^R \rangle.
\eeq
In principle, this can be achieved by choosing the multiplicative
operator renormalization
\beq \label{app5}
\left ( \begin{array}{c} \cO^R \\ \cE_i^R \end{array} \right ) =
\left ( \begin{array}{cc} \tilde Z_{\cO \cO} & \tilde Z_{\cO \cE_j} \\
\tilde Z_{\cE_i \cO} & \tilde Z_{\cE_i \cE_j} \end{array} \right )
\left ( \begin{array}{c} \cO \\ \cE_j \end{array} \right )
\eeq
in such a way that the $\langle \cE_i^R \rangle$ vanish. Plugging
Eq.\ (\ref{app5}) into Eq.\ (\ref{app4}) and using the matrix elements
(\ref{app3}), this condition (together with the finiteness of $\langle
{\cal H}_{eff}^R \rangle$) enables us to determine the renormalization constant $\tilde Z$.  Note that $\tilde
Z$ is the inverse of the usual operator renormalization constant $Z$.

However, this procedure involves of course a {\em finite}
renormalization in the $\tilde Z_{\cE_i \cO}$ in order to remove the
finite contribution $c_i \langle \cO \rangle ^{(0)}$ of the
evanescents to the matrix element (\ref{app3}). The general structure
of the renormalization constant $\tilde Z$ at $n$ loops is therefore
\beq \label{app6}
\tilde Z = \left ( \begin{array}{c|cccccc} s & \overbrace{s \; \ldots\; s}^n &  0& 0&
\ldots \\ \hline 
f \\ f & & s\\ \vdots  \end{array} \right ),
\eeq
where $s$ means ``only singular parts'' and $f$ ``also finite
parts''. Note that finite $O(\epsilon^0)$ contributions to the $b_i$
and $d_{ij}$ in Eq.\ (\ref{app3}) nevertheless yield no finite
contribution to the matrix element due to the vanishing of the
evanescents in four dimensions. This is the reason why the $\tilde Z_{\cO
\cE_j}$ and $\tilde Z_{\cE_i \cE_j}$ in (\ref{app6}) contain only
singular pieces. Furthermore, $\tilde Z_{\cO \cE_j}=0$ for $j >n$
since only the first $n$ evanescent operators contribute to the $n$ loop
matrix element $\langle \cO \rangle ^{(n)}$.

However, there is a price to pay for the finite renormalization in
$(\ref{app6})$. Namely, the finite parts of the $\tilde Z_{\cE_i \cO}$
will in general result in an additional finite contribution to the
anomalous dimension $\gamma_{\cO \cO}$ of the {\em physical} operator
$\cO$. Hence, the running and thereby the numerical value of the
Wilson coefficient $C_\cO$ in Eq.\ (\ref{app4}) is  modified in a
calculable way although the evanescent matrix elements themselves
do not contribute. This subtle effect was first observed in this
context in \cite{buras:90}.

The relation between the renormalization constant $Z=\tilde Z^{-1}$
and the corresponding anomalous dimension
\beq \label{app7}
\gamma = \tilde Z \mu {d Z \over d \mu}
\eeq
can easily be calculated at $n$ loops. From Eq.\ (\ref{app6}) it follows
for the general structure of $Z$
\beq \label{app8}
 Z = \left ( \begin{array}{c|cccccc} s & s & s & \ldots \\ \hline 
f \\ f & & s\\ \vdots  \end{array} \right ).
\eeq
Using (\ref{app6}), (\ref{app8}) and the finiteness of $\gamma$ one
obtains in the usual way
\beq \label{app9}
\gamma_{\cO \cO} = -g {d Z^{(1)}_{\cO \cO} \over dg} 
- \sum_{k=1}^n \tilde Z^{(1)}_{\cO \cE_k} g {d 
Z^{(0)}_{\cE_k \cO} \over dg},
\eeq
where we have expanded $Z$ and $\tilde Z$ in powers of $1/\epsilon$,
\bea
Z&=&Z^{(0)}+{1 \over \epsilon} Z^{(1)} + O(\epsilon^{-2}),\nonumber \\
\tilde Z &=&\tilde Z^{(0)}+{1  \over \epsilon} \tilde Z^{(1)} +
O(\epsilon^{-2}). \label{app10}
\eea
The second term in Eq.\ (\ref{app9}) would vanish without the finite
contribution due to the evanescents, so Eq.\ (\ref{app9}) indeed makes 
sense as generalization of the usual formula (\ref{2andim}).

Now one can immediately conclude from Eq.\ (\ref{app9}) that there is
no contribution to $\gamma_{\cO \cO}$ from the evanescents {\em in the
large $N_f$ limit}. This is for the following reasons: $N_f$ enters
the matrix elements (\ref{app3}) and therefore the renormalization
constants $Z$, $\tilde Z$ only at $O(\alpha_s^2)$, i.\,e.  \bea
\label{app11} Z_{\cE_k \cO} = \alpha_s A_k + \alpha_s^2 N_f B_k +
\ldots \nonumber, \\ \tilde Z_{\cO \cE_k} = \alpha_s C_k + \alpha_s^2
N_f D_k + \ldots.  \eea Multiplying these we see that the $N_f$ terms
in the evanescent contributions to the r.h.s. of Eq.\ (\ref{app9})
start with $\alpha_s^3 N_f$, so they are suppressed compared to the
$\alpha_s^3 N_f^2$ terms contained in $Z_{\cO \cO}^{(1)}$. In the
limit $N_f \rightarrow \infty$ they can therefore be neglected. Note
that this statement is only true if there are no $O(\alpha_s^0)$ terms
present in Eq.\ (\ref{app11}). In order to ensure this we defined the
evanescent operators as described in the footnote above.

\end{appendix}

\end{document}